\documentclass[prb,twocolumn,preprintnumbers,amsmath,amssymb]{revtex4}
\usepackage{bm}
\usepackage{amsmath}
\usepackage{dcolumn}
\usepackage[dvips]{graphicx}

\begin{document}
\bibliographystyle{apsrev}

\title{Stability of the spiral phase in the 
2D extended t-J model}

\author{Valeri N. Kotov}
\email{valeri.kotov@epfl.ch}
\affiliation{Institute of Theoretical Physics, Swiss Federal Institute of Technology (EPFL),\\
CH-1015 Lausanne, Switzerland}

\author{Oleg P. Sushkov}
\email{sushkov@phys.unsw.edu.au}
\affiliation{School of
Physics, University of New South Wales,\\ Sydney 2052, Australia}

\begin{abstract}
We analyze the $t-t'-t''-J$ model at low doping 
$\delta \ll 1$ by chiral perturbation theory 
and show that the (1,0) spiral state is stabilized by the presence of 
$t',t''$ above  critical values around $0.2J$, assuming  $t/J=3.1$. 
We find that the (magnon mediated) hole-hole interactions have an important effect on the
region of charge stability in the space of parameters $t',t''$, generally increasing stability,
while the stability in the magnetic sector is guaranteed by the presence of 
spin quantum fluctuations (order from disorder effect).
These conclusions are based on perturbative analysis performed up to two loops,
 with very good convergence.
\end{abstract}
\maketitle

\section{Introduction \protect\\ and summary of basic notation}

The nature of the ground state of doped quantum antiferromagnets
is a central issue for the theory of correlated electrons.
\cite{Subir} While it is well established that arbitrary small doping destroys
conventional N\'eel order, it is far less clear what is the structure of the
 emerging ground state, and how it co-exists with superconductivity.
One of the early proposals made in Ref.~\onlinecite{SS}, and later 
 explored in the context of the Hubbard and the $t-J$ models,
\cite{Kane,Auer,SS1,Mori,Sarker,Zlatko,Dupuis,Kampf,ZH,CM,SK}
 was that for small doping the collinear  N\'eel order gives way to a non-collinear spiral
 state. Energy is gained since the holes can hop easier in a spiral background. 
 However it was also established that the spiral state is itself unstable
 with respect to long wavelength density fluctuations, i.e. has a tendency towards
 phase separation, signaled by a negative charge compressibility.
The general problem of what states are susceptible to instabilities, such
 as phase separation, and what is the true ground state of the doped antiferromagnet,
 is enjoying a lot of current attention.
\cite{Kampf,WS1,VS,Pryadko,Hell,Becca,Dima} 
Some works favor, under certain conditions,
  inhomogeneous ground states, such as stripes, where the holes segregate 
 into ordered structures, \cite{WS1,VS} whereas others argue that ultimately
 the ground state is uniform. \cite{Hell,Becca} 

We have recently revisited the problem of  stability of
 the spiral state in the extended $t-t'-t''-J$ model, \cite{SK}
 and have found that the uniform spiral state is stable (at low doping) above certain 
 critical values of $t',t''$. In addition we have shown that superconductivity
 is supported in the stable region. Partial motivation for this research was
provided by the considerable body of evidence that  (incommensurate)
 magnetism co-exists with superconductivity in the LSCO family. \cite{Julien}
 Such experiments are usually interpreted in the context of stripes, 
\cite{Emery} however the (uniform) spiral scenario is quite consistent
 with the majority of data, especially in the range of doping where
 the charge order is "fluctuating", i.e. not in the ground state.
 Disorder is also expected to be important in these compounds, and can
 be readily incorporated into the spiral state. \cite{Glass}

In the present work  we use 
  chiral perturbation theory 
\cite{CP,Adler} as our main technical tool. The theory  provides a rigorous perturbative 
treatment of long-wave-length dynamics with Goldstone quasiparticles in a system with 
 strong interactions.
The starting point is the ground state of the Heisenberg model ($t-J$ model at half filling)
which incorporates all spin quantum fluctuations.
The chiral perturbation theory allows  a regular calculation of all physical
quantities in the leading  order approximation 
in powers of doping $\delta$. Subleading powers
of $\delta$ depend on the short-range dynamics and hence cannot be calculated without uncontrolled 
approximations. Therefore we cannot determine reliably 
what is the exact value of $\delta$ so that it is small enough to justify our calculations.
However  the limit  $\delta \ll 1$ justifies  the approach  parametrically.

 In our previous work  \cite{SK} we concluded that the  
spiral state is stable in a certain region of parameters $t'$ and $t''$. It was shown that the
 magnetic stability is
due to the existence of spin quantum fluctuations (order from disorder effect).
While the strong renormalization of one hole properties due to scattering with magnons 
was taken into account in Ref.~\onlinecite{SK}, the calculation 
of the compressibility at finite doping (charge response)  was made for free holes
(zero-loop approximation), and the calculation of the magnetic response was performed in 
the one-loop approximation.
In the present paper we extend our calculations up to two loops for both the charge and
the magnetic response. The results demonstrate that perturbation theory converges very well. 
The main conclusion that the magnetic stability is provided by spin quantum fluctuations 
(order from disorder effect) remains valid in two loops, however the higher corrections 
are found to have an important effect on the charge stability  region in the space of 
parameters $t',t''$, generally increasing stability.

 We start by summarizing some results and the notation of Ref.~\onlinecite{SK}
 which will be needed for our calculations.  In the S=1/2 extended $t-J$
 model, one allows  hopping 
to nearest-neighbor sites ($t$), as well as (diagonal)  next nearest-neighbors
$t'$, and next next nearest sites $t''$ on a 2D square lattice:  
\begin{eqnarray}
\label{H}
H&=&-t\sum_{\langle ij \rangle \sigma} c_{i\sigma}^{\dag}c_{j\sigma}
-t'\sum_{\langle ij_1 \rangle \sigma}
c_{i\sigma}^{\dag}c_{j_1\sigma}
-t''\sum_{\langle ij_2 \rangle \sigma}
c_{i\sigma}^{\dag}c_{j_2\sigma} \nonumber \\
&&+J \sum_{\langle ij \rangle \sigma} \left({\bf S}_i.{\bf S}_j
-\frac{1}{4}n_in_j\right).
\end{eqnarray}
The $c_{i\sigma}^{\dag}$ operators act in the space with no douple electron
 occupancy.
The number of added holes, per site, is denoted by $\delta$, and is referred
 to as doping. We also set the total number of sites $N=1$ and
 measure all the energies from now on in units of $J$ which we set to one,
 while $t$ is set to its physical value:
\begin{equation}
\label{t}
t=3.1 \ , \ \ \ J\equiv1.
\end{equation}
The same-sublattice hoppings $t',t''$ will be considered as parameters,
 having in mind that
 in the cuprates  their values are argued to be
 $t'\approx-0.8,  \ t''\approx0.7$, from fits of ARPES
 measurements. \cite{And, Kim}  

The one-hole Green's function is calculated in the self-consistent
 Born approximation (SCBA), which produces reliable results due to the
 absence of low-order vertex corrections. \cite{Kampf,SCBA1,SCBA2}
 The hole dispersion minima are located at the points  ${\bf k}_0=(\pm\pi/2,\pm\pi/2)$,
which are the centers of the four faces of the magnetic Brillouin zone (MBZ). 
In the vicinity of these points the dispersion is quadratic:
\begin{equation}
\label{disp}
\epsilon_{\bf k}\approx \frac{\beta_1}{2}k_1^2+\frac{\beta_2}{2}k_2^2, 
\end{equation}
where ${\bf k}$ is defined with respect to ${\bf k}_0$, and $k_1$ is perpendicular
 to the face of the MBZ, while $k_2$ is parallel to it.
 The quasiparticle residue near ${\bf k}_0$ is $Z_{\bf k} \approx Z_{{\bf k}_0} \equiv Z$.
By implementing the SCBA equations numerically, we have obtained the coefficients
 in the following expansion, valid in the range $-1<t'<0$, $0<t''<1$: \cite{SK}  
\begin{eqnarray}
\label{ttt}
\beta_1 \!&=& \!1.96\! +\! 1.15t'\!+\!0.06t'^2
+2.70t''+0.53t''^2+0.50t't'',\nonumber\\
\beta_2 \!&=& \!0.30\! -\! 1.33t'\!-\!0.19t'^2+2.80t''+1.06t''^2-0.14t't'',\nonumber\\
Z \!&=&\! 0.29 + 0.055t'+0.195t''.
\end{eqnarray}
Then the renormalized hole-magnon interaction is written in the long-wavelength limit,
 by introducing the operator
$\pi_{\bf q}=\alpha_{\bf q}-\beta^{\dag}_{\bf -q}$, where $\alpha_{\bf q},\beta_{\bf q}$
are the two spin waves in a two-sublattice antiferromagnet. Calling the
 two sublattices ``a" and ``b" and introducing the corresponding (renormalized) hole 
operators
$h_{{\bf k}a},h_{{\bf k}b}$, we obtain the 
Hamiltonian describing the interactions between holes and spin waves:
\begin{equation}
\label{hsw}
H_{h,sw}=\sum_{\bf k,q}M_{\bf q} \left(
h_{{\bf k+q}a}^{\dag}h_{{\bf k}b}\pi_{\bf q}+
\mbox{h.c.}\right).
\end{equation}
For simplicity from now on we refer  to the renormalized (dressed) holes
simply as holes.
The  vertex in (\ref{hsw}) is ($q_1$ is perpendicular to the MBZ face):
\begin{equation}
\label{M}
 M_{\bf q} =
-2^{7/4}Zt\frac{q_1}{\sqrt{|{\bf q}|}}. 
\end{equation}
We stress that $M_{\bf q} \to 0$ at $q\to 0$ in accordance with the general Adler's
relation. \cite{Adler}
In a spiral state the spins deviate from the N\'eel order,
 leading to lowering of the ground state energy and creation of a gap
in the hole spectrum. The spiral pitch $Q$ can be directed along the
(1,1) or (1,0) direction of the lattice and  is proportional
 to doping $Q \sim \delta $. 
The new hole operators $\psi, \varphi$, and the corresponding 
energies are:
\begin{equation}
\label{ops}
\left(\begin{array}{c} 
\psi_{{\bf k}}^{\dag} \\ \\ \varphi_{{\bf k}}^{\dag}
\end{array} \right)
=\frac{1}{\sqrt{2}}\left(
h_{{\bf k}a}^{\dag}\!\mp\!e^{-i\mu}h_{{\bf k}b}^{\dag}\right); \
\left(\begin{array}{c}
\epsilon_{{\bf k}}^{\psi} \\ \\
\epsilon_{{\bf k}}^{\varphi}
\end{array}  \right)
=\epsilon_{{\bf k}} \mp \frac{\Delta}{2}.
\end{equation}
The spiral is a coplanar state with spins lying in a fixed plane.
The arbitrary phase $\mu$ is related to the orientation of this
plane,
 which, as we will see below, does not appear
 in any physical observables. 
The operators $\psi_{{\bf k}}$ and $\varphi_{{\bf k}}$ describe spinless fermions,
and due to (\ref{ops}) the $\varphi$
fermion band is completely empty. The gap and the Fermi energy are:  
\begin{equation}
\label{gap}
\Delta = \frac{2Z^2t^2}{\pi\rho_s\sqrt{\beta_1\beta_2}} \ \epsilon_F \ ,
 \ \ \epsilon_F= \frac{1}{N_p}2\pi\sqrt{\beta_1\beta_2} \ \delta.
\end{equation}
Here $\rho_s =Z_{\rho}/4, \ Z_{\rho}=0.72$ (S=1/2) is the renormalized spin stiffness
 of the undoped antiferromagnet, and  $N_p$ stands for the number of full pockets
 enclosed by the Fermi surface. For the (1,0) spiral $N_p=2$, while  
 for the 45$^{\circ}$ spiral (1,1) there is only one pocket  $N_p=1$. \cite{SK}
Notice that except for the band splitting (\ref{ops}) which absorbs the soft mode
of the doped Heisenberg antiferromagnet we assume that the dispersion (\ref{disp}) 
remains rigid under doping.

The rest of the paper is organized as follows.
 In Section II we  analyze the charge stability
 of the spiral states, while Section III is devoted to the stability in the magnetic sector.
The dispersion variation due to doping is discussed in Section IV.
Section V contains our conclusions.

\section{Charge response and charge stability}

We now proceed to calculate the charge compressibility $\chi$, defined
as $\chi^{-1}=\partial^{2}E/\partial\delta^{2}$, where $E$ is the ground state
energy.
\begin{figure}
\centering
\includegraphics[height=180pt,keepaspectratio=true]{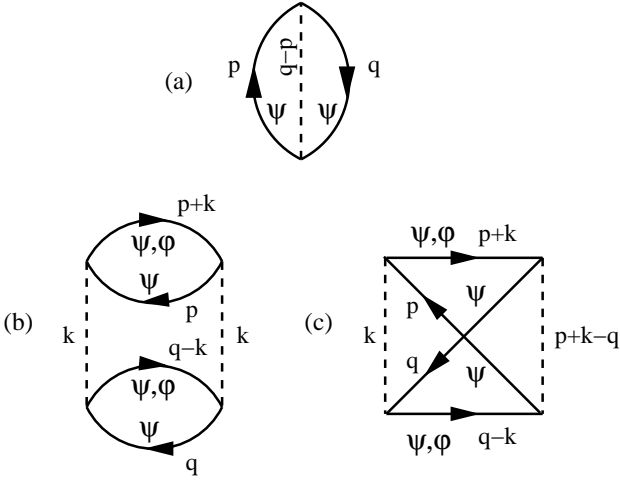}
\caption{(a) First order interaction energy. Dashed line represents
the vertex (\ref{Gamma}). (b) Second order direct energy. (c.) Second order exchange.
The labels $\psi$ and $\varphi$ show which type of fermion is present in the intermediate state. }
\label{Fig1}
\end{figure}
\noindent
The two simplest contributions to the energy come from the Fermi motion of holes and
from the energy gain due to the spiral formation. These two contributions are: 
\cite{SK}
\begin{equation}
\label{e0}
E^{(0)}= \frac{1}{N_p}\left(\pi\sqrt{\beta_1\beta_2}
-\frac{Z^2t^2}{\rho_s} \right)\delta^{2} \ . 
\end{equation}
The superscript $(0)$ in the energy shows that  interaction 
effects have not been taken into account (no loops).
We  note that the terms that  are not quadratic in $\delta$ are not explicitly
written, i.e. we have subtracted from the energy
the constant and linear terms: $E_{AF} + {\mbox{const.}}\ \delta$.

The holes interact via spin wave exchange. Treating the hole-magnon interaction
(\ref{hsw}) in the second order of perturbation theory 
(the $h_a,h_b$ operators have to be expressed via the $\psi,\varphi$
operators from (\ref{ops})) we obtain the effective low-energy hole-hole 
interaction vertex:
\begin{equation}
\label{Gamma}
\Gamma_{\bf q} \approx -\frac{M_{\bf q}^2}{\omega_{\bf q}}=
-8Z^2t^2 \frac{q_1^2}{|{\bf q}|^2}.
\end{equation}
Here we have neglected both the retardation effects as well as the
 hole energies in the denominators. This approximation should
work well in the limit of low doping $\delta \ll 1$ since 
typical hole energies $\epsilon_{\bf k} \sim \epsilon_F \sim \delta$,
while the magnon energy $\omega_{\bf q} \sim |{\bf q}| \sim k_F 
\sim \sqrt{\delta} \gg \delta$, for $\delta \ll 1$.
Notice also that $Zt \approx 1$ and consequently the interaction is strong even in the
long-wave-length limit.

The one-loop correction to the energy is given by the exchange diagram in Fig.~1(a).
The corresponding expression reads
\begin{equation}
\label{exch}
E^{(1)} = -\frac{1}{2} \sum_{{\bf p},{\bf k}}\Gamma_{\bf p-k} n_{{\bf p}}n_{{\bf k}},
\end{equation}
where the summations are within one hole pocket enclosed by the elliptic
Fermi surface, 
$n_{{\bf k}}= \theta \left(\epsilon_{F} - \epsilon_{\bf k} \right)$.
The direct term, proportional to $\Gamma_{\bf q =0}$ is set to zero,
by keeping in mind the presence of a small non-zero frequency in the 
denominator of (\ref{Gamma}).
Notice the overall sign (+) in Eq.~(\ref{exch})
which is opposite to the sign appearing for an electron gas with Coulomb interactions.
This is due to the fact that the interaction (\ref{Gamma}) is attractive.
Introducing the  rescaled variables: $\tilde p_{1}=p_{1} \sqrt{\beta_{1}/(2\epsilon_F)}$,
 $\tilde p_{2}=p_{2} \sqrt{\beta_{2}/(2\epsilon_F)}$, we have explicitly:
\begin{equation}
\label{exch1}
E^{(1)} = \frac{Z^2t^2\epsilon_{F}^{2}}{\beta_{1}\beta_{2}\pi^4} \int f_{{\bf \tilde p}-{\bf \tilde k}} n_{{\bf \tilde p}}
n_{{\bf \tilde k}} d^{2}\tilde k d^{2}\tilde p,
\end{equation}
where the normalized distribution is:
$n_{{\bf \tilde k}}= \theta (1-|{\bf \tilde k}|)$,
and we define the function:
\begin{equation}
f_{\bf \tilde k} = \frac{\tilde k_{1}^{2}}{\tilde k_{1}^{2}+(\beta_1/\beta_2) \tilde k_{2}^{2}}.
\end{equation}
An explicit analytical evaluation of  $E^{(1)}$ is not possible, but the numerical integration
in (\ref{exch1}) does not cause any problems.

\begin{figure}
\centering
\includegraphics[height=90pt,keepaspectratio=true]{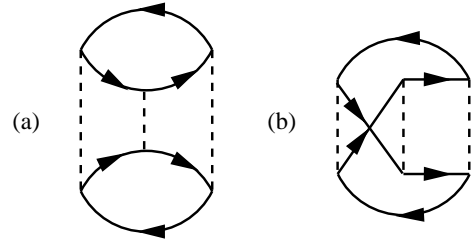}
\caption{Third order ladder diagrams. (a) Direct. (b.) Exchange.}
\label{Fig2}
\end{figure}
\noindent

The two-loop corrections to the energy are given by the diagrams in Fig.~1(b,c),
which contain both $\psi-\psi$ and $\psi-\varphi$ scatterings. 
The first type of contribution produces:
\begin{eqnarray}
E^{(2)}_{\psi-\psi}&=&\frac{1}{2} \sum_{{\bf p},{\bf k},{\bf q}}\frac{\Gamma_{\bf k}^2 - 
\Gamma_{\bf k}\Gamma_{\bf p+k+q}}{{\epsilon^{\psi}_{\bf p}+\epsilon^{\psi}_{\bf q}-\epsilon^{\psi}_{\bf p+k}
-\epsilon^{\psi}_{\bf q+k}}} \nonumber \\
&&\times
n_{{\bf p}}n_{{\bf q}}(1-n_{{\bf p+k}})(1-n_{{\bf q+k}}),
\end{eqnarray}
while the second one is:
\begin{equation}
E^{(2)}_{\psi-\varphi}=\frac{1}{2} \sum_{{\bf p},{\bf k},{\bf q}}\frac{\Gamma_{\bf k}^2 -
\Gamma_{\bf k}\Gamma_{\bf p+k+q}}{{\epsilon^{\psi}_{\bf p}+\epsilon^{\psi}_{\bf q}-\epsilon^{\varphi}_{\bf p+k}
-\epsilon^{\varphi}_{\bf q+k}}}n_{{\bf p}}n_{{\bf q}}.
\end{equation}
Here both the direct and exchange terms are added up, as shown in Fig.~1(b,c).
In the rescaled variables we have the  form convenient for numerical integration:
\begin{eqnarray}
\label{exch2}
&E^{(2)}_{\psi-\psi}&=
-\frac{2Z^4t^4 \epsilon_{F}^{2}}{(\beta_{1}\beta_{2})^{3/2} \pi^{6}}
\int [f_{\bf \tilde k}^{2} - f_{\bf \tilde k}f_{{\bf \tilde k}+{\bf \tilde p}+{\bf \tilde q}} ]
n_{{\bf \tilde p}}n_{{\bf \tilde q}} \nonumber \\
&&
\times(1-n_{{\bf \tilde p+ \tilde k}})
(1-n_{{\bf \tilde q+ \tilde k}}) \frac{ d^{2}\tilde k d^{2}\tilde p d^{2}\tilde q}
{[{\bf \tilde k}.({\bf \tilde k + \tilde p+ \tilde q})]} \nonumber \\
&E^{(2)}_{\psi-\varphi}&= -\frac{2Z^4t^4 \epsilon_{F}^{2}}{(\beta_{1}\beta_{2})^{3/2} \pi^{6}}
\int
[f_{\bf \tilde k}^{2} - f_{\bf \tilde k}f_{{\bf \tilde k}+{\bf \tilde p}+{\bf \tilde q}} ] \nonumber \\
&&
\times n_{{\bf \tilde p}}n_{{\bf \tilde q}} \frac{ d^{2}\tilde k d^{2}\tilde p d^{2}\tilde q}
{[{\bf \tilde k}.({\bf \tilde k + \tilde p+ \tilde q})+\Delta/\epsilon_F]}  
\end{eqnarray}
Notice that the energy corrections (\ref{exch2}) do not contain infrared or ultraviolet
divergences in contrast to the 3D electron gas with Coulomb interaction. \cite{GB}
In fact each of the  two-loop diagrams in Fig.~\ref{Fig1} contains an ultraviolet
logarithmic divergence,  $\ln \Lambda$, where $\Lambda \sim \frac{1}{\sqrt{\delta}}
\gg 1$ is an ultraviolet cut-off.
However    the divergences are canceled out between the direct
and the exchange diagrams, since the fermions $\psi, \varphi$
 are spinless. 

In the next, third  order of perturbation theory the only
diagrams containing ultraviolet divergences are the "ladder" ones, shown in  Fig.~\ref{Fig2}.
Both diagrams separately are proportional to $\ln^2 \Lambda$, and, once
again, 
 the divergences are canceled out exactly when the  direct and exchange terms
are added together.
By examining also  higher orders we find quite generally that the perturbative
 expansion behaves well.
The absence of  divergences is
due to the dimensionality of the problem as well as the Goldstone nature of
the mediator of the interaction.

\begin{figure}
\centering
\vspace{0.6cm}
\hspace{-0.6cm}
\includegraphics[height=150pt,keepaspectratio=true]{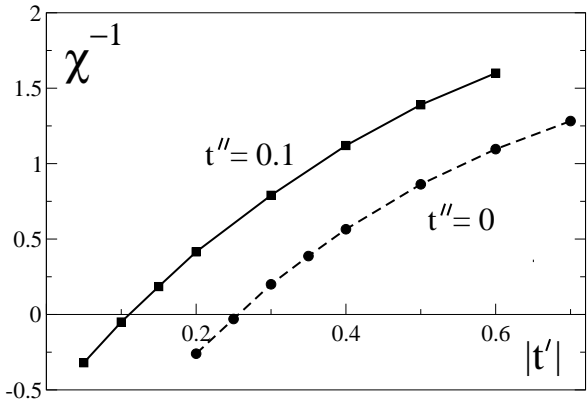}
\caption{The inverse compressibility $\chi^{-1}$, as  defined 
\newline
in  Eq.~(\ref{e11}), calculated for $t'<0$.}
\label{Fig3}
\end{figure}
\noindent

We have verified that the hole-magnon vertex only acts within a single pocket, 
i.e. holes from different pockets do not interact. Thus, the total 
ground state energy including one and two-loop corrections is:
\begin{equation}
\label{int}
E=E^{(0)}+ N_p[E^{(1)}+E^{(2)}_{\psi-\psi}+E^{(2)}_{\psi-\varphi}] \ .
\end{equation}
All terms in this expression  are proportional to $\delta ^2$ and all
the terms scale as $1/N_p$ since expressions (\ref{exch1}) and (\ref{exch2})
contain $\epsilon_F^2$. Therefore,
defining $\chi^{-1}$ to be the inverse compressibility of the (1,0) spiral state,
i.e. $E_{(1,0)}=\frac{1}{2}\chi^{-1} \delta^{2}$, we have:
\begin{equation}
\label{e11}
E_{(1,1)}=2E_{(1,0)} = \chi^{-1} \delta^{2} \ .  
\end{equation}
Thus whenever the (1,0) state is lower in energy $E_{(1,0)}<E_{(1,1)}$,
 it is stable $\chi^{-1}>0$, whereas the (1,1) state is always unstable,
because if $E_{(1,1)}<E_{(1,0)}$ then $E_{(1,1)}<0$ and hence
$\chi^{-1}_{(1,1)}<0$.

\begin{figure}
\vspace{0.6cm}
\centering
\includegraphics[height=175pt,keepaspectratio=true]{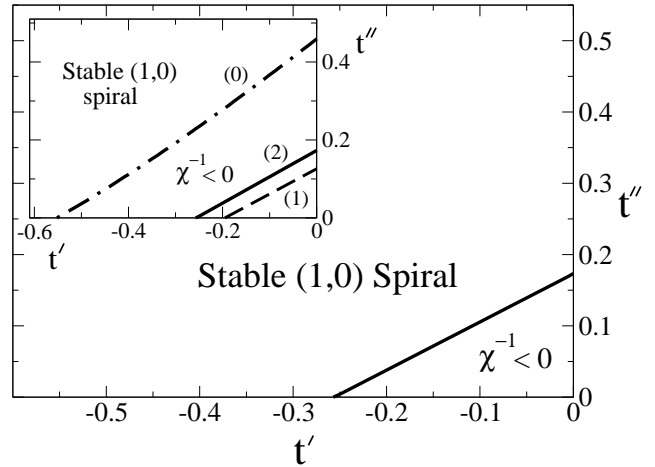}
\caption{Region of stability of the spiral state in the
 $t-t'-t''-J$ model at $t=3.1$ (throughout the work we take $J=1$). 
The unstable region, below the critical
 line, is marked as $\chi^{-1}<0$. Inset: Stability boundaries calculated in 
different orders of perturbation theory. The dot-dashed line shows the result 
obtained in Ref.~\onlinecite{SK} without the interaction corrections (zero loops).
 The dashed line is the stability boundary calculated in the one-loop approximation 
($E^{(0)}$ and $E^{(1)}$ in Eq. (\ref{int})).
The solid line, which is the same in the inset and in the main figure, 
shows the stability line calculated in the two-loop approximation.
}
\label{Fig4}
\end{figure}
\noindent

In Eqs.~(\ref{int},\ref{e11}) we have neglected the contact
 interaction energy
$E^{\mbox{c}}$, arising from the last (density-density) term in (\ref{H}), proportional
 to $J=1$. In fact one could imagine adding an additional nearest neighbor
 Coulomb repulsion energy $V>0$ to (\ref{H}) of the form $H_{V}= V \sum_{\langle ij \rangle} n_in_j$.
 Then the total contact energy to lowest order is: $E_{(1,1)}^{\mbox{c}}=0$, 
$E_{(1,0)}^{\mbox{c}}= (V-1/2)Z^{2}\delta^{2}$. The presence of the $V$ term generally
 increases $\chi^{-1}$; however as long as $V \sim 1$, the contact energy's
 contribution to the total energy is extremely small, since all the terms in
  Eq.~(\ref{int}) contain  powers of the dominant scale $t\approx3$. 

We have found numerically that the relative importance of each successive order of perturbation theory
 decreases roughly by a factor of 3. For example, we estimate the first order
 with respect to the kinetic (Fermi motion) energy:
$N_pE^{(1)}/E_{F} \approx 1/3$. The second-order
contribution relative to the first order is also approximately:
 $|E^{(2)}_{\psi-\psi}+E^{(2)}_{\psi-\varphi}|/E^{(1)} \approx 1/3$.
 We have therefore stopped at the second order of perturbation theory and
 expect  the contribution of the next orders  to be negligible.

The inverse compressibility $\chi^{-1}$ is shown in Fig.~3 as a function of $t'$ for 
two values of $t''$. When the point $\chi^{-1}=0$ is reached, the system becomes
unstable. The stability analysis leads us to the phase diagram 
of Fig.~4: the spiral phase is stable above the solid line.
In Fig.~4(inset) we show the  boundaries of stability calculated in different orders
of perturbation theory. The dot-dashed line  is the zero loop result,
 taking into account $E^{(0)}$ only (no interaction effects),  while  the  
dashed line is the 
boundary calculated in the one-loop approximation ($E^{(0)}$ and $E^{(1)}$ in Eq.~(\ref{int})).
Finally, the  solid line is 
 the stability border calculated in the two-loop approximation.
The insert demonstrates the  good convergence of the perturbation theory.
It is interesting that account of the interaction corrections (one and two-loop corrections)
enlarges the region of stability in the space of parameters $t',t''$.

\section{Magnon stability}

 We now proceed to verify that the  region which is 
stable with respect to density fluctuations ($\chi^{-1}>0$), shown in Fig.~4, is also stable
 in the magnetic sector. The starting point is the spin-wave Green's function:\cite{SK}
\begin{equation}
\label{D}
D_{n}(\omega,{\bf q}) \! = \!
\frac{2\omega_{\bf q}\left[\omega^2-\omega_{\bf q}^2-2\omega_{\bf q}P_n(\omega,{\bf q})
\right]}
{\left[\omega^2-\omega_{\bf q}^2-2\omega_{\bf q}P_n(\omega,{\bf q})\right]^2  \!-  \!
4\omega_{\bf q}^2\left|P_a(\omega,{\bf q})\right|^2}, 
\end{equation}
where $\omega_{\bf q}=\sqrt{2}|{\bf q}|, \ |{\bf q}|\ll 1$,
 is the magnon energy, and $P_n,P_a$ are the normal and
 anomalous polarization operators which can be calculated perturbatively
 for the interaction Eq.~(\ref{hsw}). The above  Green's  function is the normal
 one (the anomalous  Green's  function $D_{a}(\omega,{\bf q})$ has the same
 denominator).

The magnetic stability requires that all the poles of (\ref{D}) are at
 positive $\omega^2$, i.e. no imaginary poles exist.
The one loop, first order contributions to $P_n,P_a$ were already calculated
 in Ref.~\onlinecite{SK}, and we now calculate the second order,
 to match the order in the ground state energy.
 Due to the fact that in the spiral state the fermion operators
 on the two sublattices are mixed via Eq.~(\ref{ops}),
 corrections to the vertex appear in  lowest order.
 For example the second order diagrams for the normal part $P^{(2)}_{n}$
 are shown in Fig.~5(a,b,c).
\begin{figure}
\centering
\includegraphics[height=265pt,keepaspectratio=true]{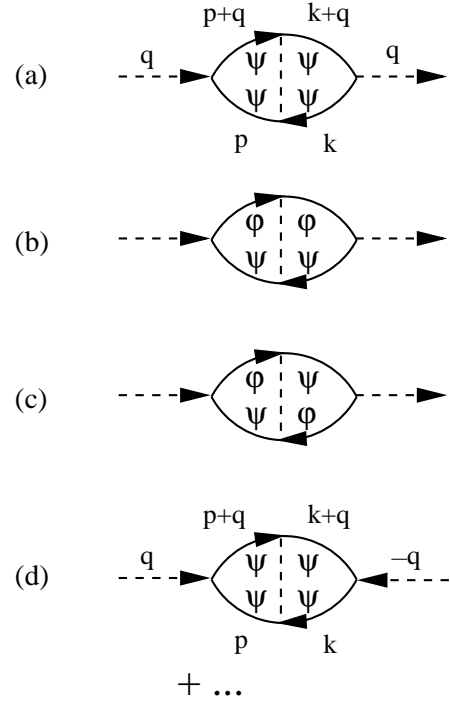}
\caption{(a,b,c) Two-loop, second order diagrams contributing to the
 normal polarization operator $P_{n}$. (d) Two-loop contributions
to the anomalous polarization $P_{a}$ (only first diagram of several is explicitly shown).}
\label{Fig5}
\end{figure}
\noindent
After evaluating these diagrams we obtain:
\begin{eqnarray}
\label{P}
P^{(2)}_{n}(0,{\bf q})\!\! &=&\!\! \frac{M_{{\bf q}}^{2}}{4} \! \sum_{{\bf p}, {\bf k}, N_p  }
\Gamma_{{\bf p}-{\bf k}} \left( \frac{(n_{{\bf k}} -  n_{{\bf k} +
 {\bf q}})(n_{{\bf p}}- n_{{\bf p} +{\bf q}})}{(\epsilon^{\psi}_{{\bf k}}-
\epsilon^{\psi}_{{\bf k}+{\bf q}})(\epsilon^{\psi}_{{\bf p}}-\epsilon^{\psi}_{{\bf p}+{\bf q}})}
\right. \nonumber \\
&&+ \frac{2n_{{\bf k}} n_{{\bf p}
+{\bf q}}}{(\epsilon^{\psi}_{{\bf k}}-\epsilon^{\varphi}_{{\bf k}+{\bf q}})
(\epsilon^{\psi}_{{\bf p}+{\bf q}}-\epsilon^{\varphi}_{{\bf p}})} \nonumber\\
&&-\left. \frac{2n_{{\bf k}} n_{{\bf p}
}}{(\epsilon^{\psi}_{{\bf k}}-\epsilon^{\varphi}_{{\bf k}+{\bf q}})
(\epsilon^{\psi}_{{\bf p}}-\epsilon^{\varphi}_{{\bf p}+{\bf q}})}
\right).
\end{eqnarray}
For the stability analysis one needs only the long wavelength, i.e.
  $|{\bf q}| \rightarrow 0$ limit of (\ref{P}). In this limit the second
 and third terms (Fig.~5(b,c)) in  (\ref{P}) cancel out,
leaving only the first term, Fig.~5(a). The corresponding diagram for
 the anomalous part (Fig.~5(d)) is: 
\begin{equation}
P^{(2)}_{a}(0,{\bf q})= -e^{-2i\mu}P^{(2)}_{n}(0,{\bf q}), \ \ |{\bf q}| \rightarrow 0.
\end{equation}

Adding to these results the one-loop results, $P_{n,a}^{(1)}$ from Ref.~\onlinecite{SK},
we obtain the polarization operators up to two loops
$P_{n,a}=P_{n,a}^{(1)}+P_{n,a}^{(2)}$.
For $|{\bf q}| \rightarrow 0$ we have explicitly:
\begin{equation}
P_{n}(0,{\bf q}) \! = \! -\frac{\sqrt{2} Z^2t^2}{\pi \sqrt{\beta_{1}\beta_{2}}}
\left(1+\frac{2\epsilon_{F}}{\Delta}\right) |{\bf q}| +
 \frac{4\sqrt{2}I Z^4t^4}{\pi^2 \beta_{1}\beta_{2}} |{\bf q}|,
\end{equation}
\begin{equation}
P_{a}(0,{\bf q}) e^{2i\mu} = 
\frac{\sqrt{2} Z^2t^2}{\pi \sqrt{\beta_{1}\beta_{2}}}
\left(1-\frac{2\epsilon_{F}}{\Delta}\right) |{\bf q}|-
\frac{4\sqrt{2}I Z^4t^4}{\pi^2 \beta_{1}\beta_{2}} |{\bf q}|
\end{equation}
\noindent
The quantity $I$ is defined as:
\begin{equation}
I\!  = \! \int_{0}^{2 \pi} \! \!\frac{dxdy}{(2\pi)^2}\frac{(\cos x-\cos y)^2}{
(\cos x\!- \! \cos y)^2 \! + \! (\beta_1/\beta_2) (\sin x\!- \! \sin y)^2}.
\end{equation}

The stability criterion reads:
\begin{equation}
\label{stab1}
\left[\omega_{\bf q}^2+2\omega_{\bf q}P_n(0,{\bf q})\right]^2
> 4\omega_{\bf q}^2\left|P_a(0,{\bf q})\right|^2,
\end{equation}
which, for the (1,0) spiral state, is equivalent to
the following inequality, in terms of the microscopic parameters:

\begin{eqnarray}
\label{stab2}
\lefteqn{\left|1-\frac{2Z^2t^2}{\pi\sqrt{\beta_1\beta_2}} -2\rho_s
+ \frac{8IZ^4t^4}{\pi^2\beta_1\beta_2} \right| >} 
\; \; \; \; \; \; \; \; \;  \nonumber\\
&&\left| \frac{2Z^2t^2}{\pi\sqrt{\beta_1\beta_2}} - 2\rho_s
-\frac{8IZ^4t^4}{\pi^2\beta_1\beta_2} \right|.\ 
\end{eqnarray}
We have verified numerically that inside the region marked ``Stable''
 in Fig.~4, the expression inside the 
absolute value sign on the left hand side of (\ref{stab2}) is positive,
 whereas the one on  the right  is negative. This means that 
  (\ref{stab2}) is equivalent to the equation 
\begin{equation}
\label{sc}
1-4\rho_{s} > 0 \ .
\end{equation}
This is  the same condition as the one obtained  \cite{SK} in the 
one-loop approximation. This condition is always satisfied since
$1-4\rho_{s}=1-Z_{\rho}$, and $Z_{\rho}=0.72$ due to spin quantum
fluctuations renormalization in the undoped S=1/2 antiferromagnet. 
The physical meaning of the ``quantum stabilization'' is quite simple.
Indeed, treating the spins semiclassically we would have $\rho_s=1/4$ and hence a 
marginal ground state. \cite{SS1,CM,SK} The spin quantum fluctuations
reduce the spin stiffness and hence make the energy of the spiral state lower,
see Eq.~(\ref{e0}). This is certainly only 
 the intuitive physical picture, while the regular proof is the one presented above,
leading to Eqs.~(\ref{stab2},\ref{sc}).

\section{Self-energy corrections and their influence on stability}

The last issue we would like to address is the variation of the hole 
dispersion with doping and its influence on the phase
 diagram of Fig.~4. While the renormalization of the one-hole properties
at zero doping has already been taken into account via the  SCBA, Eq.~(\ref{ttt}), at finite doping
the many-body
 hole-hole interaction corrections originating from the vertex  (\ref{Gamma}) have to be calculated.

The leading   self-energy diagram, contributing 
to the hole dispersion at finite doping is shown in Fig.~6.
The corresponding expression is
\begin{equation}
\label{se1}
\Sigma_{\bf k}^{(1)}=
-\sum_{\bf q}\Gamma_{{\bf k}-{\bf q}}n_{\bf q}=
8Z^2t^2\sum_{\bf q}\frac{(k_1-q_1)^2}{|{\bf k}-{\bf q}|^2} n_{\bf q} \ .
\end{equation}
The self-energy is the same for $\psi$ and $\varphi$ fermions and therefore it influences
the dispersion (\ref{disp}), but does not influence the gap $\Delta$ between the bands.
The one-loop self-energy depends only on the momentum of the particle.
We remind the reader that the one-loop self-energy is closely related to the one-loop 
correction to the total energy, namely by cutting a fermionic line in
diagram Fig.~\ref{Fig1}(a) one obtains the self-energy Fig.~6.
This means that the self-energy effect has already been taken into account
in the calculation of the total energy performed in Section II.
Nevertheless it is interesting to see explicitly how the dispersion changes.
The integral in (\ref{se1}) can be easily evaluated numerically at any
point of the phase diagram Fig.~\ref{Fig4}. However it is more instructive
to consider the case of isotropic dispersion $\beta_1=\beta_2=\beta$
where the integration can be performed analytically. 
In this case
\begin{eqnarray}
\label{se2}
\Sigma_{\bf k}^{(1)}&=&
\frac{2Z^2t^2\epsilon_F}{\pi\beta}\left[1+ \Omega_k (k_1^2-k_2^2)\right]\ ,\nonumber \\ 
\Omega_k&=&\left\{\begin{array}{l} 
\frac{1}{2k_F^2} \ ,\hspace{1.7cm} \mbox{for} \ \ k < k_F\\ 
\frac{1}{k^2}\left(1-\frac{k_F^2}{2k^2}\right) \ ,  \ \   \mbox{for} \ \ k > k_F \ .
\end{array} \right.
\end{eqnarray}
With account of the self-energy the dispersion (\ref{disp}) should be replaced to
$\epsilon_{\bf k} \to \epsilon_{\bf k}+\Sigma_{\bf k}$. Thus effectively we have
a deformation
of the Fermi surface
\begin{equation}
\beta_1=\beta+\frac{Z^2t^2}{\pi} \ , \ \ 
\beta_2=\beta-\frac{Z^2t^2}{\pi} \ .
\end{equation}
\begin{figure}
\centering
\includegraphics[height=75pt,keepaspectratio=true]{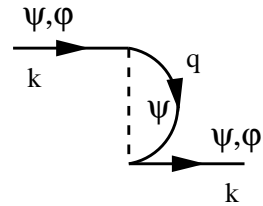}
\caption{One-loop self-energy correction. Dashed line represents
the vertex (\ref{Gamma}). The correction is the same for $\psi$ and $\varphi$ fermions.
 }
\label{Fig6}
\end{figure}
\noindent
Notice that the deformation does not depend on the 
doping $\delta$. On the other hand numerically this effect turns out to be
 rather weak.
 For the purpose of an estimate we take $t'=-0.8, \ t''=0.7$,
corresponding to the  physical values of  these  parameters. \cite{And,Kim}
At this point $\beta_1=2.95$, $\beta_2=3.80$, $Zt=1.19$, hence the Fermi surface
deformation is $\Delta\beta/\beta\approx Z^2t^2/(\pi\sqrt{\beta_1\beta_2})\approx 0.13$.

Finally, we  can summarize our findings: (i.) The leading order
 self-energy correction produces a weak, doping independent spectrum renormalization.
 (ii.) The charge stability analysis of Section II performed up to (two-loop) order
 $(Zt)^{4}$ is not influenced, at that order, by the change in the spectrum.
Taking the latter effect into account produces higher order diagrams and is expected
 to be a very small correction, due to the previous point (i).
(iii.) The magnon stability analysis of Section III is not influenced
 by the inclusion of  (\ref{se2}), since the
 the density of states 
$\propto 1/\sqrt{\beta_1\beta_2}$ cancels out  in the stability criterion (\ref{sc}).

\section{Conclusions}

We have studied the stability of the spiral
 states in the  $t-t'-t''-J$ model at low doping $\delta \ll 1$.  
The stability was studied both in the charge and in the spin sectors, 
and takes into account interaction effects  by including diagrams up to 
two loops. Our main result is that 
relatively small values of $t',t''$ are sufficient to stabilize the (1,0) spiral
 state, as shown in Fig.~4. For example for $t''=0$, the critical
 value of $t'$ is  $|t_{c}'/J| \approx 0.25$, or in units of $t$,   $|t_{c}'| \approx 0.08 t$
(for $t/J=3.1$).
 The unstable region for small $t',t''$ is a good candidate for
an inhomogeneous ground state of some sort, e.g. a striped one.
 Our results are also consistent with DMRG studies \cite{WS2} of the extended $t-J$
model predicting a transition  from  stripes at $t'=0$ to
 a homogeneous ground state around $|t'/t| \approx 0.1$.
Parameters for real cuprates correspond to the top left corner of the phase diagram Fig.~4
where the (1,0) spiral state is stable.

 We have  shown that the interaction effects (higher loop diagrams) should be taken into account
 and they tend to increase the region of spiral stability in the space
 of parameters $t',t''$.  The effective coupling constant governing the perturbation
 theory is $g=(Zt)^2/\pi \approx 0.3$. Therefore each successive order is several
 times weaker than the previous one, and thus we expect our results, calculated up
 to second order, to be quite reliable. The good convergence is demonstrated by the inset in 
Fig.~4 where we show the critical lines calculated in zero, one and two-loop
orders. The magnetic stability of the spiral state is guaranteed by spin quantum fluctuations
(order from disorder effect) -  this conclusion remains valid in the two-loop approximation.

Our calculations are based on the chiral perturbation theory, therefore they are
reliable only in the limit $\delta \ll 1$, whereas the  finite doping renormalizations 
of the various Green's functions and vertices become more and more substantial as doping 
increases, leading essentially to an untreatable problem.

\begin{acknowledgments}
V.N.K. acknowledges the support of  the Swiss National Fund.
\end{acknowledgments}

\end{document}